\documentclass[12pt,thmsa]{article}
\usepackage{sw20aip}

\input{tcilatex}
\begin{document}

\title{Some linear differential expressions for an electron scattering problem in a
field of the one-dimentional arbitrary potential }
\author{David M. Sedrakian$^{*}$,Ashot Zh. Khachatrian$^{**}$ \\
$^{*}$Department of Physics, Yerevan State University, Manookian 1,\\
375049 Yerevan, Armenia \\
$^{**}$Department of Physics, Armenian State Engineering University,\\
Terian 3, 375046 Yerevan, Armenia}
\maketitle

\begin{abstract}
The linear system of differential equations for determination of
transmission and reflection amplytudes of scattered electron in the field of
one dimensional arbitrary potential is obtained. It is shown that in general
the scattering problem can be reduced to the Caushy problem for the
stationary Shrodinger equation. An explicit functional for a dependance of
transmission amplitude from scattered field is found.
\end{abstract}

\section{Introduction}

There have been many studies for investigation quantum particle propagation
in the one-dimensional nonregular systems \cite{Lif.}. It is well known that
this problem has approaches in many fields of physics , particularly, it has
important role for the transport theory of different perturbations in the
linear quaziperiodical, disordered and random systems. An electron
scattering problem in the field of the one-dimensional arbitrary potential
is one of the key problem of such theories. Therefore to have a method for
calculation transmission amplitude (TA) and reflection amplitude (RA) of an
electron for one-dimensional arbitrary shape potential is actual.

This problem is the problem which is included in quantum mechanics
textbooks. It is well known that the problem of determination TA and RA is
reduced to the solution of the Shrodinger equation for wave function $(\hbar
^2=2m_0=1)$

\begin{equation}
\frac{d^2\Psi }{dx^2}+(E-V(x))\Psi =0  \label{eq1}
\end{equation}
Here $V(x)$ is an arbitrary limited function which tends to zero, when $%
x\rightarrow \pm \infty .$

Asymptotic behavior of the solution for $\Psi $ is written in the form of

\begin{equation}
\Psi (x)=e^{ik_0x}+R\,e^{-\,ik_0x}\text{,}\,\,\,\,\,\,when\text{\ }%
x\rightarrow -\,\infty \text{,}  \label{eq2}
\end{equation}

\begin{equation}
\Psi (x)=Te^{ik_0x}\text{,}\,\,\,\,\,\,when\ x\rightarrow +\,\infty \text{,}
\label{eq3}
\end{equation}
where $T$ and $R$ are electron TA and RA for potential $V(x)$ and $k_0=\sqrt{%
E}$. In this approach to find $T$ and $R$ one have to solve the wave
equation (\ref{eq1}) with conditions (\ref{eq2}),(\ref{eq3}).

Using the phase function method one can directly calculate $T$ and $R$ \cite
{Col.,Bab.}. In this method scattering amplitudes are considered as a
function of coordinate $x$, so that $T(x)$ and $R(x)$ correspond to TA and
RA for the part of potential closed in the interval $(x,+\infty )$. Then,
for functions $T(x)$ and $R(x)$ it was obtained a system of nonlinear
equations. Particularly, the equation for RA is an ordinary Recatty equation:

\begin{equation}
\frac{dR(x)}{dx}=\frac{iV(x)}{2k_0}(e^{ik_0x}+R\,e^{-\,ik_0x})^2\text{,}
\label{eq4}
\end{equation}

\[
R(\infty )=0\text{,} 
\]
where $R(x)$ is RA for potential 
\begin{equation}
U(x)=V(x^{^{\prime }})\theta (x-x^{^{\prime }})\text{,}  \label{eq5}
\end{equation}
$\theta (x)\;$is step function. The equation (\ref{eq4}) is well studied,
but its analytical solution for given potential is not always known.

The aim of this paper's to show, that in contrast to (\ref{eq4}), with the
help of convenient choice of unknown function as a combination $T(x)$ and $%
R(x),$the problem of determination TA and RA is reduced to the problem of
solution of the system of linear differential equations . We obtained this
result from the consideration of the problem of an electron propagation in
the one-dimensional chain with potential

\begin{equation}
V(x)=\sum\limits_{n=1}^NV_n(x-x_n)\text{,}  \label{eq6}
\end{equation}
where $V(x-x_n)$ is an individual potential of the chain located near the
point $x_{n\text{ }}$. It is suggested as well, that individual potentials
do not have general points. Using the linear recurrent expressions for TA
and RA for potential (\ref{eq6}) and approximating an arbitrary function
with the help of rectangular potentials , we obtain the linear differential
equations for determination $T$ and $R$.

Some results of this work connected to the plane wave propagation through
one-dimensional arbitrary linear medium were published in the paper \cite
{Sed.}.

This work consist from 7 sections. In (Sec. 2) the recurrent equations for
TA and RA in one-dimensional chain is found. In (Sec. 3) the system of
lineal differential equations for determination scattering amplitudes for a
potential of an arbitrary shape is obtained. Further, in (Sec. 4) we show
that the scattering problem can be formulated as a Caushy problem for
Srodinger equation. In (Sec. 5) the explicit functional for TA from
scattered potential is found. In conclusion the advantage and connection of
suggested method to well known approach are discussed.

\section{Recurrent relations for $T_N$ and $R_N$}

Let us consider the problem of determination of \thinspace transmission
amplitude $T_N$ and reflection amplitude $R_N$ for $\,$the potential (\ref
{eq6}). It is well known that this problem is reduced to calculation of the $%
N$ matrixes product \cite{Erd.,Azb.}:

\begin{equation}
\left( 
\begin{array}{c}
T_N \\ 
0
\end{array}
\right) =\prod_N^{n=1}\left( 
\begin{array}{cc}
1/t_n^{*} & -r_n^{*}/t_n^{*} \\ 
-r_n/t_n & 1/t_n
\end{array}
\right) \left( 
\begin{array}{c}
1 \\ 
R_N
\end{array}
\right) \text{,}  \label{eq7}
\end{equation}
where $t_n$ and $r_{n\text{ }}$are TA and RA of the individual potential $%
V_n(x-x_n)$. When $V_n(x-x_n)$ are rectangular potentials with arbitrary
widths $d_n$ and magnitudes of potential $V_n$, the scattering amplitudes $%
t_n$ and $r_{n\text{ }}$are given by the well known formulas \cite{Blo.}: 
\begin{equation}
t_n^{-1}=\exp ik_0d_n\;\left\{ \cos k_nd_n-i\frac{k_n^2+k_0^2}{2k_nk_0}\sin
k_nd_n\right\} ,  \label{eq8}
\end{equation}
\begin{equation}
r_n/t_n=i\exp i2k_0x_n\;\frac{k_n^2-k_0^2}{2k_nk_0}\sin k_nd_n.  \label{eq9}
\end{equation}
where $k_n=\sqrt{E-V_n}$, $x_n$ is a coordinate of the middle point of the
potential.

Let us show, that the problem of calculation of the $N$-th two-order
matrixes product (\ref{eq7}) is equivalent to solution of some linear
recurrent equation. Let us denote

\begin{equation}
\left( 
\begin{array}{ll}
\,\,\,\,\,\,\,1/T_{N-1}^{*} & -R_{N-1}^{*}/T_{N-1}^{*} \\ 
-R_{N-1}/T_{N-1} & \,\,\,\,\,1/T_{N-1}
\end{array}
\right) =\prod_{N-1}^2\left( 
\begin{array}{cc}
1/t_n^{*} & -r_n^{*}/t_n^{*} \\ 
-r_n/t_n & 1/t_n
\end{array}
\right) \text{,}  \label{eq10}
\end{equation}
then, as it is clear from (\ref{eq10}), $T_{N-1}$and $R_{N-1}$are TA and RA
of potential (\ref{eq6}) in which the first th and $N$th individual
potentials are absent..

From (\ref{eq7}) and (\ref{eq10}) for quantities $S_N=1/T_N$ and $\overline{S%
}_N=R_N/T_N$ the following recurrent relations can be obtained;$\smallskip $%
\begin{equation}
S_N=1/t_Nt_1\,S_{N-1}+r_Nr_1^{*}/t_Nt_1^{*}\,S_{N-1}^{*}+r_1^{*}/t_Nt_1^{*}%
\overline{S}_{N-1}+r_N/t_Nt_1\,\overline{S}_{N-1}^{*},  \label{eq11}
\end{equation}

\begin{equation}
\overline{S}_N=1/t_Nt_1^{*}\,\overline{S}_{N-1}+r_Nr_1/t_Nt_1\,\overline{S}%
_{N-1}^{*}+r_1/t_Nt_1S_{N-1}+r_N/t_Nt_1^{*}\,S_{N-1}^{*},  \label{eq12}
\end{equation}

The recurrent equations(\ref{eq11}), (\ref{eq12}) will be used to solve the
problem (\ref{eq1})-(\ref{eq2}) for an arbitrary potential $V(x)$.

\section{The system of linear differential equations for $T(x_1,x_2)$ and $%
R(x_1,x_2)$}

\smallskip

Let us introduce functions $S(x_1,x_2)=1/T(x_1,x_2)$ and $\overline{S}%
(x_1,x_2)=R(x_1,x_2)/T(x_1,x_2)$, where $T(x_1,x_2)$ and $R(x_1,x_2)$ are TA
and RA for the part of potential $V(x)$ , which is included between the
points $x_1$ and $x_2$, so if $x_1<$ $x_2$ we have

\begin{equation}
U(x,x_1,x_2)=V(x)\theta (x-x_1)\theta (x_2-x)\text{,}  \label{eq13}
\end{equation}

Then the function $U(x,x_1-\Delta x_1,x_2+\Delta x_2)$, when $\Delta x_1$and 
$\Delta x_2$ are small enough, can be represented as the function $%
U(x,x_1,x_2)$ with two rectangular potentials added to it. One of the
rectangular potentials added to the left side will be characterized by the
value of potential $V(x_1)$ and the width $\Delta x_1$, second one added to
the right side will be characterized by the value of potential $V(x_2)$ and
the width $\Delta x_2.$

From (\ref{eq8}), (\ref{eq9}) for TA and RA for infinitely narrow
rectangular potential $(d_n<<1)$ we have 
\begin{equation}
t_n^{-1}=1+\frac{iV_nd_n}{2k_0}\,\,\,\,,\,\,r_n/t_n=-\,\frac{iV_nd_n}{2k_0}%
\exp i2k_0x_n\,\,\,\,\,\,  \label{eq14}
\end{equation}

Let us substitute in (\ref{eq11}), (\ref{eq12}) $S_N$ as $S(x_1-\Delta
x_1,x_2+\Delta x_2)$ and $\overline{S}_N$ as $\overline{S}(x_1-\Delta
x_1,x_2+\Delta x_2)$, $S_{N-1}$ as $S(x_1,x_2)$ , $\overline{S}_{N-1}$ as$%
\overline{\text{ }S}(x_1,x_2)$ and expand the obtained expressions in a
series in infinitely small quantities $\Delta x_1,\Delta x_2$. Then, taking
into account (\ref{eq14}), it is possible to obtain the system of following
linear differential equations ;

\begin{equation}
\frac{\partial S}{\partial x_1}=-\frac{iV(x_1)}{2k_0}S-\frac{iV(x_1)}{2k_0}%
\exp \{-2ik_0x\}\overline{S}\text{,}  \label{eq15}
\end{equation}

\begin{equation}
\frac{\partial \overline{S}}{\partial x_1}=\frac{iV(x_1)}{2k_0}\overline{S}+%
\frac{iV(x_1)}{2k_0}\exp \{2ik_0x\}S\text{,}  \label{eq16}
\end{equation}

\begin{equation}
\frac{\partial S}{\partial x_2}=\frac{iV(x_2)}{2k_0}S-\frac{iV(x_2)}{2k_0}%
\exp \{2ik_0x\}\overline{S}^{*}\text{,}  \label{eq17}
\end{equation}
\begin{equation}
\frac{\partial \overline{S}}{\partial x_2}=\frac{iV(x_2)}{2k_0}\overline{S}-%
\frac{iV(x_2)}{2k_0}\exp \{2ik_0x\}S^{*}\text{,}  \label{eq18}
\end{equation}

The solutions of the system (\ref{eq15})-(\ref{eq18}) have to be satisfied
the following initial conditions

\begin{equation}
S(x_1,x_2)|_{x_1=x_2}=1,\;\,\,\,\overline{S}(x_1,x_2)|_{x_1=x_2}=0
\label{eq19}
\end{equation}

Let us now show that the law of current density conservation follows from (%
\ref{eq15})-(\ref{eq18}):

\begin{equation}
\left| S(x_1,x_2)\right| ^2-\,\left| \overline{S}(x_1,x_2)\right| ^2=\left|
T(x_1,x_2)\right| ^2+\,\left| R(x_1,x_2)\right| ^2=1  \label{eq20}
\end{equation}
for any $x_1$ and $x_2.$ Differentiating (\ref{eq20}) on $x_1$ and $x_2$ we
get

\begin{equation}
S^{*}\frac{\partial S}{\partial x_1}+S\frac{\partial S^{*}}{\partial x_1}-%
\overline{S}^{*}\frac{\partial \overline{S}}{\partial x_1}-\overline{S}\frac{%
\partial \overline{S}^{*}}{\partial x_1}=0  \label{eq21}
\end{equation}

\begin{equation}
S^{*}\frac{\partial S}{\partial x_2}+S\frac{\partial S^{*}}{\partial x_2}-%
\overline{S}^{*}\frac{\partial \overline{S}}{\partial x_2}-\overline{S}\frac{%
\partial \overline{S}^{*}}{\partial x_2}=0  \label{eq22}
\end{equation}
Using equations (\ref{eq15})-(\ref{eq18}) and (\ref{eq21}),(\ref{eq22}) it
is easy to see the correctness of condition (\ref{eq20}).

Let us consider (\ref{eq15}),(\ref{eq16}) for fixed $x_2$ and variable $x_1$%
. Denoting $S(x_1,x_2)\equiv P(x)$ and $\overline{S}(x_1,x_2)\equiv 
\overline{P}(x)$ for functions $P(x)$ and $\overline{P}(x)$ the following
differential equations are obtained;

\begin{equation}
\frac{dP}{dx}=-\frac{iV(x)}{2k_0}P-\frac{iV(x)}{2k_0}\exp \{-2ik_0x\}%
\overline{P}\text{,}  \label{eq23}
\end{equation}

\begin{equation}
\frac{d\overline{P}}{dx}=\frac{iV(x)}{2k_0}\overline{P}+\frac{iV(x)}{2k_0}%
\exp \{2ik_0x\}P\text{,}  \label{eq24}
\end{equation}

The initial conditions for (\ref{eq23})-(\ref{eq24}) are

\begin{equation}
P(\infty )=1\text{ and \thinspace \thinspace \thinspace \thinspace }%
\overline{P}(\infty )=0  \label{eq25}
\end{equation}

The equations (\ref{eq23})-(\ref{eq24}) describe the dependance of
scattering parameters $T(x)$ and $R(x)$ on $x$ for potential

\[
U(x)=V(y)\theta (x-y). 
\]

In the case of the fixed $x_1$ and variable $x_2$ for the functions $%
S(x_1,x_2)\equiv D(x)$ and $\overline{S}(x_1,x_2)\equiv \overline{D}(x)$ the
following equations are obtained from (\ref{eq17}),(\ref{eq18});

\begin{equation}
\frac{dD}{dx}=\frac{iV(x)}{2k_0}D-\frac{iV(x)}{2k_0}\exp \{2ik_0x\}\overline{%
D}\text{,}  \label{eq26}
\end{equation}

\begin{equation}
\frac{d\overline{D}}{dx}=-\frac{iV(x)}{2k_0}\overline{D}+\frac{iV(x)}{2k_0}%
\exp \{-2ik_0x\}D  \label{eq27}
\end{equation}
The initial conditions for (\ref{eq26}),(\ref{eq27}) are

\begin{equation}
D(-\infty )=1\text{ and \thinspace \thinspace \thinspace \thinspace }%
\overline{D}(-\infty )=0  \label{eq28}
\end{equation}

The equations (\ref{eq26}),(\ref{eq27}) describe the dependance of
scattering parameters $T(x)$ and $R(x)$ on $x$ for the potential

\[
U(x)=V(y)\theta (y-x) 
\]

So the problem of the determination $T$ and $R$ is reduced in general to the
problem of solution of the system differential equations (\ref{eq23})-(\ref
{eq24}) or (\ref{eq26}),(\ref{eq27}) with initial conditions (\ref{eq25}) or
(\ref{eq28}).

\smallskip

\section{ The problem of a Cauchy for a scattering parameters.}

\smallskip

In this section we will show that, using equations (\ref{eq23}),(\ref{eq24})
or (\ref{eq26}),(\ref{eq27}) it is possible to get the two linear equations
for combinations of functions $P(x)$ , $\overline{P}(x)$ or $D(x)$ , $%
\overline{D}(x)$, one of which is the Shrodinger equation, another one is
the primary first order differential equation. Let us present (\ref{eq23})-(%
\ref{eq24}) in the following form:

\begin{equation}
\exp \{ik_0x\}\frac{dP}{dx}=-\frac{iV(x)}{2k_0}\left[ P\exp \{ik_0x\}+%
\overline{P}\exp \{-ik_0x\}\right] \text{,}  \label{eq29}
\end{equation}

\begin{equation}
\exp \{-ik_0x\}\frac{d\overline{P}}{dx}=\frac{iV(x)}{2k_0}\left[ P\exp
\{ik_0x\}+\overline{P}\exp \{-ik_0x\}\right] \text{.}  \label{eq30}
\end{equation}

Let us introduce the functions $F_1$and $\overline{F}_1:$

\begin{equation}
P\exp \{ik_0x\}=F_1\text{ and }\overline{P}\exp \{-ik_0x\}=\overline{F}_1
\label{eq31}
\end{equation}

and write (\ref{eq29})-(\ref{eq30}) in the form of

\begin{equation}
\frac{dF_1}{dx}-ik_0F_1=-\frac{iV}{2k_0}\{F_1+\overline{F}_1\}\text{,}
\label{eq32}
\end{equation}

\begin{equation}
\frac{dF_1}{dx}+ik_0F_1=\frac{iV}{2k_0}\{F_1+\overline{F}_1\}\text{.}
\label{eq33}
\end{equation}

From these equations it is easy to show that for quantities $F_1+\overline{F}%
_1=L_1$ and $F_1-\overline{F}_1=Q_1$we have equations

\begin{equation}
\left[ \frac{d^2}{dx^2}+E-V(x)\right] L_1(x)=0,Q_1=-\frac i{k_0}\frac{dL_1}{%
dx}\text{.}  \label{eq34}
\end{equation}

Introducing quantities $D\exp \{-ik_0x\}=F_2$ and $\overline{D}\exp
\{ik_0x\}=\overline{F}_2$, and using equations (\ref{eq26})-(\ref{eq27}) for
the functions $F_2-\overline{F}_2=L_2$ and $F_2+\overline{F}_2=Q_2$ the
following equations are received:

\begin{equation}
\frac d{dx}Q_2=-\frac i{k_0}\left( k_0^2-V(x)\right) L_2\text{,}
\label{eq34a}
\end{equation}

\begin{equation}
\frac{dL_2}{dx}=-ik_0Q_2\text{.}  \label{eq35a}
\end{equation}
Excluding the function $Q_2$ from equation (\ref{eq34a}), we get

\begin{equation}
\left[ \frac{d^2}{dx^2}+E-V(x)\right] L_2(x)=0,Q_2=\frac i{k_0}\frac{dL_2}{dx%
}\text{.}  \label{eq35}
\end{equation}

Let us now apply the equation (\ref{eq35}) to the problem of determination
of TA and RA for an arbitrary potential $V(x)$ with finite size (the
potential $V(x)$ equals to zero outside of some interval $a\leq x\leq b$)..

Inserting $L_2\equiv L$ , $F_2\equiv F$ and $\overline{F}_2\equiv \overline{F%
}$ , then from (\ref{eq35}) we get

\begin{equation}
F=\frac 12\left( \frac i{k_0}\frac{dL}{dx}+L\right) ,\,\,\overline{F}=\frac
12\left( \frac i{k_0}\frac{dL}{dx}-L\right)  \label{eq36}
\end{equation}
and

\begin{equation}
\left[ \frac{d^2}{dx^2}+E-V(x)\right] L(x)=0\text{.}  \label{eq37}
\end{equation}
The boundary conditions for function $L(x)$ at the point $x=a$ are

\begin{equation}
L(a)=\exp \{-ik_0a\},(dL/dx)|_{x=a}=-ik_0\exp \{-ik_0a\}\text{.}
\label{eq38}
\end{equation}
Let us seek solution of (\ref{eq37}) in the form

\begin{equation}
L(x)=\exp \{-ik_0a\}\left( H_1(x)-ik_0H_2(x)\right) .  \label{eq39}
\end{equation}
Then the real functions $H_1$ and $H_2$ satisfy the equation (\ref{eq37})
with boundary conditions

\begin{equation}
H_1(a)=1,(dH_1/dx)|_{x=a}=0\,\,and\,H_2(a)=0,(dH_2/dx)|_{x=a}=1.
\label{eq.40}
\end{equation}
Substituting the solution (\ref{eq39}) into (\ref{eq36}) and taking into
account the connection between $F,\,\overline{F}$ and $D,\,\overline{D}$ we
obtain

\begin{equation}
\frac 1T=\frac 12\exp \{ik_0d\}\left[ H_1+\frac{dH_2}{dx}-ik_0H_2+\frac
i{k_0}\frac{dH_1}{dx}\right] \text{,}  \label{eq41}
\end{equation}

\begin{equation}
\frac RT=\frac 12\exp \{i2k_0x_0\}\left[ -H_1+\frac{dH_2}{dx}-ik_0H_2-\frac
i{k_0}\frac{dH_1}{dx}\right] \text{,}  \label{eq42}
\end{equation}
where $x_0=(b+a)/2$ and $d=b-a$.

Thus, we showed, that the solution of the problem (\ref{eq1})-(\ref{eq3}),
i.e. the problem of determination RA and TA for an arbitrary potential, is
reduced to a Cauchy problem for the Srodinger equation(\ref{eq37}).

In the end of this section, using equations (\ref{eq34a}),(\ref{eq35a}), we
will bring the differential equations for the transmission coefficient $%
\left| T\right| ^2=\left| D\right| ^{-2}$ and reflection coefficient $\left|
R\right| ^2=\left| \overline{D}\right| ^2/\left| D\right| ^2$. The result
which we obtained can be represented in the following way. If the unknown
functions $\left| D\right| ^2$and $\left| \overline{D}\right| ^2$ are
expressed by the function $M$ as

\begin{equation}
\left| D\right| ^2=\frac 12\left( \frac M{2k_0^2}+1\right) ,\,\left| 
\overline{D}\right| ^2=\frac 12\left( \frac M{2k_0^2}-1\right) \text{,}
\label{eq42a}
\end{equation}
then the function $M(x)$ are connected with the function $G(x)$ by
differential equation:

\begin{equation}
\frac{dM}{dx}=V(x)\frac{dG}{dx}\text{,}  \label{eq43a}
\end{equation}
and function $G(x)$ is the solution of the equation:

\begin{equation}
\frac{d^3G}{dx^3}+4(E-V(x))\frac{dG}{dx}-2\frac{dV}{dx}G=0\text{,}
\label{eq44a}
\end{equation}
with initial conditions

\begin{equation}
G(a)=1,\;(dG/dx)|_{x=a}=0,\,(d^2G/dx^2)|_{x=a}=2V(a)\text{.}  \label{eq45a}
\end{equation}

\section{Functional $T\left[ V(x)\right] $}

In this section we bring an explicit form of the functional $T\left[
V(x)\right] $. To do it, let us assume that in (\ref{eq11}) and (\ref{eq12}) 
$t_1=0,\;r_1=1$. It means, that we add an rectangular potential only from
the right hand side of the chain. For this case, inserting $S_N\equiv
D_N,\,S_N\equiv D_N^{*}$ , one can write (\ref{eq11}), (\ref{eq12}) in the
form

\begin{equation}
D_N=\frac{r_N}{t_N}\overline{D}_{N-1}+\frac 1{t_N}D_{N-1}\text{,}
\label{eq43}
\end{equation}

\begin{equation}
\overline{D}_N=\frac{r_N^{*}}{t_N^{*}}D_{N-1}+\frac 1{t_N^{*}}D_{N-1}\text{.}
\label{eq44}
\end{equation}

Note, that here $N$ is a variable. Let us represent (\ref{eq43}), (\ref{eq44}%
) in a more convenient form. Excluding $\overline{D}_{N-1}$ from the first
equation and $D_{N-1}$ from the second one, we get two independent equations
for $D_N$ and $\overline{D}_N$:

\begin{equation}
D_N=A_ND_{N-1}-B_ND_{N-2},\,N\geq 2  \label{eq45}
\end{equation}

\begin{equation}
\overline{D}_N=A_N^{*}\overline{D}_{N-1}-B_N^{*}\overline{D}_{N-2},\,N\geq 2
\label{eq46}
\end{equation}
where $A_N=\frac 1{t_N}+B_N/t_{N-1}^{*}$ and $B_N=r_Nt_N/t_Nr_N$.

Let us apply the obtained result (\ref{eq45}) for the potential consisted
from infinitely narrow rectangular potentials (\ref{eq14}). In this case $%
A_N $ and $B_N$ are

\begin{equation}
A_N=1+B_N+\frac{iV_Nd_N}{2k_0}\left[ 1-\exp \{i2k_0(x_N-x_{N-1})\}\right] 
\text{,}  \label{eq47}
\end{equation}

\begin{equation}
B_N=\frac{V_Nd_N}{V_{N-1}d_{N-1}}\exp \{i2k_0(x_N-x_{N-1})\}\text{.}
\label{eq48}
\end{equation}

Representing $D_N$ in the form

\begin{equation}
D_N=1+\sigma _N+\sigma _{N-1}+\cdot \cdot \cdot +\sigma _1\text{,}
\label{eq49}
\end{equation}
where

\[
\sigma _N=\frac{iV_Nd_N}{2k_0}\{1+f_{N,N-1}\sigma _{N-1}+f_{N,N-2}\sigma
_{N-2}+\cdot \cdot \cdot +f_{N,1}\sigma _1\} 
\]
and $f_{N,n}=\exp \{i2k_0(x_N-x_n)\}$, $\sigma _1=\frac{iV_1d_1}{2k_0}$, for 
$D_N=T_N^{-1}$ the following expression is found

\begin{equation}
T_N^{-1}=1+\sum_{p=1}^N\,\,\sum_{1\leq n_1<\cdot \cdot \cdot <n_p}^N\frac{%
iV_{n_1}d_{_{n_1}}}{2k_0}\cdot \cdot \cdot \frac{iV_{n_p}d_{n_p}}{2k_0}%
\prod_{l=1}^{p-1}\left[ 1-\exp \{i2k_0(x_{n_{l+1}}-x_{n_l})\}\right] \text{.}
\label{eq50}
\end{equation}

Putting $V_n$ as $V(x_n)$ and tending $N\rightarrow \infty $ and $\max
d_n\rightarrow 0$ it is possible from (\ref{eq50}) to obtain the following
series:

\begin{equation}
T^{-1}=1+\sum_{n=1}^\infty W_n\text{,}  \label{eq51}
\end{equation}
where

\begin{equation}
W_n=\stackrel{\infty }{\stackunder{x_{n-1}}{\int }}\stackrel{x_{n-1}}{%
\stackunder{x_{n-2}}{\int }}\cdot \cdot \cdot \stackunder{-\infty }{%
\stackrel{x_{n_1}}{\int }}\frac{iV(x_1)}{2k_0}\cdot \cdot \cdot \frac{iV(x_n)%
}{2k_0}\prod_{l=1}^{n-1}\left[ 1-\exp \{i2k_0(x_{l+1}-x_l\}\right] dx_1\cdot
\cdot \cdot dx_n  \label{eq52}
\end{equation}

The expression (\ref{eq51}) are the explicit expression for functional $%
T\left[ V(x)\right] $.

\section{Conclusion}

The important result of this work is the suggestion, that for scattering
problem instead of solving the Srodinger equation for $\Psi $ (\ref{eq1})-(%
\ref{eq3}),we can consider the system of linear equations for $T^{-1}$and $%
R/T$, i.e. to consider equations (\ref{eq23}),(\ref{eq24}) or (\ref{eq26}),(%
\ref{eq27}).

Particularly, from these equations one can derivide the Recatty equation (%
\ref{eq4}), for $R(x)$. This equation nonlinear and its analytical solutions
are known for a limited class of potentials.

Here we have shown that the problem of determination of the functions $T(x)$
and $R(x)$ is reduced to solution of the linear equation for function $L$,
which is some given combination from $T(x)$ and $R(x)$. It is important to
note, that this equation coincides with the well studied Srodinger equation.

The another important result is the differential equations for transmission
and reflection coefficients for an arbitrary potentials $V(x)$. Since the
value of function $G$ and their derivations are known in the initial point
of the potential, the problem can be formulated as Caussy problem.

Supposing, that potential $V(x)=cons^{\prime }t$, when $a\leq x\leq b$ and $%
V(x)=0$, when $x<a,\,x>b$, the well known result for transmission
coefficient for rectangular one can obtain from (\ref{eq42a})-(\ref{eq45a}).
Indeed, solving equations (\ref{eq43a})-(\ref{eq45a}) when $V(x)=V_0$, for
function $M(x)$ we get

\begin{equation}
\frac M{2k_0^2}=\frac{V_0^2}{2k_0k^2}\sin ^2kx+1\text{.}  \label{eq53}
\end{equation}
Putting (\ref{eq53}) in (\ref{eq42a}) we obtain the following expression for
transmission coefficient of the rectungular potential\cite{Blo.}:

\begin{equation}
\left| T\right| ^2=\frac{4k_0^2k^2}{(k_0^2-k^2)^2\sin ^2kx+4k_0^2k^2}
\label{eq54}
\end{equation}
where $k=\sqrt{E-V_0}$.

We would like to thank D. Badalyan for useful discussions.

\end{document}